# Computational diffraction reveals long-range strains, disorder and crystalline domains in atomic scale simulations

A. Boulle[1], A. Chartier[2], A. Debelle[3], X. Jin[1,3], J. -P. Crocombette[4]

[1]Institut de Recherche sur les Céramiques, CNRS UMR 7315, Limoges, France

[2]CEA, DEN, DPC, SCCME, Gif-Sur-Yvette, France.

[3]Laboratoire de Physique des 2 Infinis Irène Joliot-Curie, Université Paris-Saclay, CNRS/IN2P3, UMR 9012, Université Paris-Saclay, Orsay, France.

[4]CEA, DEN, SRMP, Gif-sur-Yvette, France.

**Abstract**

Atomic scale simulations are a key element of modern science in that they allow to understand, and even predict, complex physical or chemical phenomena on the basis of the fundamental laws of nature. Among the different existing atomic scale simulation approaches, molecular dynamics (MD) has imposed itself as the method of choice to model the behavior of the structure of materials under the action of external stimuli, say temperature, strain or stress, irradiation, etc. Despite the widespread use of MD in condensed matter science, some basic material characteristics remain difficult to determine. This is for instance the case of the long-range strain tensor in heavily disordered materials, or the quantification of rotated crystalline domains lacking clearly defined boundaries. In this work, we introduce computational diffraction as a fast and reliable structural characterization tool of atomic scale simulation cells. As compared to usual direct-space methods, computational diffraction operates in the reciprocal-space and is therefore highly sensitive to long-range spatial correlations. With the example of defective $UO_2$, it is demonstrated that the *homogeneous* strain tensor, the *heterogeneous* strain tensor, the disorder, as well as rotated crystallites are straightforwardly and unambiguously determined. Computational diffraction can be applied to any type of atomic scale simulation and can be performed in real time, in parallel with other analysis tools. In experimental workflows, diffraction and microscopy are almost systematically used together in order to benefit from their complementarity. Computational diffraction, used together with computational microscopy, can potentially play a major role in the future of atomic scale simulations.

# 1. Introduction

Atomic scale simulations play a central role in modern scientific research in that they, in principle, allow to compute and predict the properties of materials using the very building blocks of matter, *viz.* atoms (Tadmor & Miller, 2011). A key aspect of such approaches relies on the ability to derive long-range, continuum-like parameters from a discrete system of particles (Costanzo *et al.*, 2005). From a strictly scientific point of view, this is a fairly old problem, that, for instance, led to the introduction of the virial theorem (Clausius, 1870). The constant increase in computational power observed these last decades, combined with the availability of efficient simulation software, are giving atomic scale simulations an ever increasing importance in all fields of science, and in particular in Materials Science.

An issue common to most materials science studies is the understanding of the evolution of the crystallographic structure of materials under the action of external stimuli, say temperature, stress or strain, laser or ion irradiation, etc. In this context, molecular dynamics (MD) simulations have emerged as the method of choice to tackle that type of issues (Nordlund & Djurabekova, 2014; Krasheninnikov & Nordlund, 2010). Efficient algorithms have been developed to characterize the local structure of materials in multi-million atoms simulation cells, such as the Voronoi analysis, bond angle analysis, common neighbor analysis, dislocation extraction algorithm, etc. (Stukowski, 2012; Stukowski *et al.*, 2012) (Reference (Stukowski, 2012) provides a review of existing theories and algorithms). These methods are particularly efficient to derive structural parameters such as the local strain, the atoms coordination, etc. Although the process of analyzing such large simulation cells can in principle be automated, thereby enabling so-called *in situ* computational microscopy (Zepeda-Ruiz *et al.*, 2017), it often remains a semi-automated process requiring manual fine tuning by expert scientists. This human-centered workflow is a time-consuming process that, *de facto,* constitutes a bottleneck when several hundreds of simulation data sets can be generated in the course of a few hours. Besides, structural parameters of primary importance, such as the long-range strain tensor or the number and size of crystalline domains are ill-defined in some situations or, at least, not straightforward to determine. For instance, a large number of studies have been dedicated to the determination of strain from atomic scale simulations in the last decades (Zimmerman *et al.,* 2009; Mott *et al.,* 1992; Gullett *et al.,* 2008; Stukowski & Arsenlis, 2012; Stukowski *et al.,* 2009; Zhang *et al.,* 2015; Xiong *et al.,* 2019). In these studies the local deformation tensor is obtained from the evolution of the coordinates of the atoms constituting a coordination shell between any two time steps of a MD trajectory ("kinematical" approach). This approach becomes problematic in highly disordered regions of the cell where the coordination shells are incomplete, like in the close vicinity of surfaces, grain boundaries or dislocation cores. This is especially true for nanostructured materials or heavily damaged crystals, as those encountered during ion irradiation where the level of disorder can bring materials into an amorphous state (Boulle & Debelle, 2016). Another example where current approaches fail is the detection of crystalline domains which, roughly speaking, rely on the detection of bond lengths or bond angles (or the associated strain tensor components) exceeding a pre-defined threshold value. This method can become an issue when the transition from one domain to another is continuous, *i.e.* without abrupt interfaces or boundaries. This situation illustrated in Fig. 1, which corresponds to the disordered $UO_2$ MD cells investigated in this work. In this figure, we selected four disorder levels, here expressed in dpU (*i.e.* displacements per U atoms, a parameter quantifying the fraction of displaced atoms, see section 3). For low levels of disorder (b), the cell remains single-crystalline with visible defects. For high levels of disorder (c and d), crystalline domains are clearly distinguished but it is not straightforward to count those domains or estimate their size as there are no clearly defined grain boundaries.

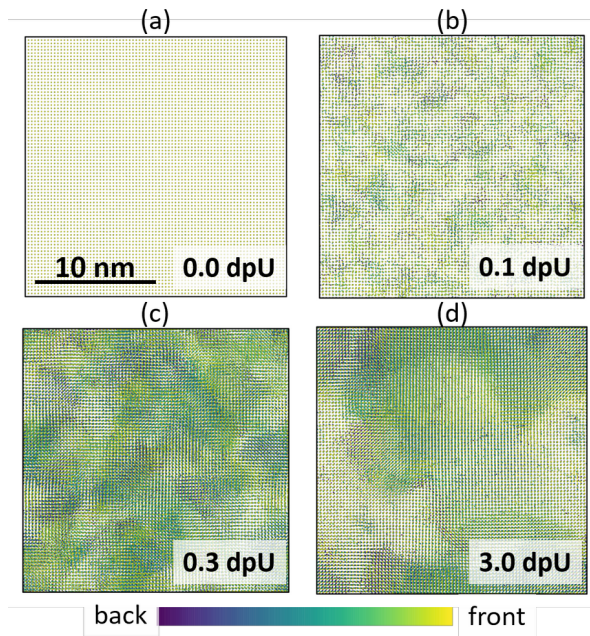

*Fig. 1. Snapshots of an $UO_2$ MD cell at different levels of induced disorder (expressed in displacement per uranium, dpU). For visualization purposes, only U atoms are displayed and the thickness normal to the figure is 5 nm.*

As far as strain is concerned, a possible workaround is to compute the "lattice strain" derived by the examination of the peak positions in a virtual X-ray diffraction (XRD) measurement (Stukowski *et al.,* 2009; Xiong *et al.,* 2019), *i.e.* in the reciprocal space. A virtual XRD pattern can be straightforwardly generated using the Debye scattering equation (DSE) (Warren, 1969). Coupling the DSE with MD simulations is of common practice for the determination of the structure of nanostructured materials (for a comprehensive review, see (Gelisio & Scardi, 2016) and references therein). Conversely, the DSE remains only marginally used for the determination of strain and disorder in damaged materials, like those obtained by laser (Lin & Zhigilei, 2006) or ion (Chartier *et al.*, 2018; Soulié *et al.*, 2016; Debelle *et al.*, 2014) irradiation, for instance. There are, moreover, several caveats for the use of the DSE to derive strain and disorder in MD cells.

Firstly, an explicit assumption underlying the DSE is that the interatomic vectors are allowed to take all orientations in space with equal probability, which implicitly implies randomly oriented polycrystalline materials or amorphous forms of matter (Warren, 1969). The resulting intensity distribution corresponds to a spherical projection of the reciprocal space onto a single radius of the sphere, *i.e.* a powder-like diffraction pattern. As a result of this spherical integration, the possibility of detecting individual crystalline domains is inevitably lost. Secondly, the peak overlap inherent in that type of signals, might hinder an accurate evaluation of the peak positions and intensities and consequently an accurate determination of the strain. In addition, all quantities derived from such an analysis are orientation-averaged, which prohibits the detection of spatial anisotropies and all information regarding rigid rotations are lost in the process. Finally, although this issue is scarcely addressed in the literature (Stukowski *et al.*, 2009; Xiong *et al.*, 2019), it is established that the local strain, directly determined from the MD cells, does not contain any information regarding the long-range correlations of the atomic displacements, contrarily to the lattice strain derived from the DSE. As a consequence, both calculations yield different values, which leads to an indeterminacy as to which measurement represents the actual state of strain of the MD cell.

In the present work, we solve all of the above-mentioned issues. Our computational diffraction approach is based on a reciprocal-space analysis of the MD cells, and it can efficiently complement existing algorithms both on a qualitative level (*i.e.* for visualization purposes) or on a quantitative level. Firstly, computational diffraction allows for a complete determination of the homogeneous strain tensor (*i.e* long-range strain) and of the heterogeneous strain tensor (*i.e.* the root-mean-squared (rms) fluctuations of the strain), as well as the determination of the orientation-dependent atomic disorder. Secondly, it makes the detection and quantification of individual crystalline domains a straightforward task. Moreover, we prove that both the reciprocal-space and the real-space determinations (that is, directly from the atomic coordinates) yield the same results if long-range correlations are correctly taken into account. Doing so, we provide a fast and self-consistent approach for the determination of crystalline domains, strain and disorder in atomic scale simulation cells. The validity of our computational diffraction approach is tested on $UO_2$ MD cells in which we simulated ion-irradiation induced disorder and for which we have a good understanding of the defect structure and evolution (Chartier *et al.*, 2016; Jin *et al.*, 2020).

**2. Theoretical background**

Two-dimensional diffraction, also referred to as reciprocal space mapping, is a widespread experimental technique to analyze strain and various types of defects in epitaxial films or single crystals (Holý *et al.*, 1999). Reciprocal space maps (RSMs) are obtained by the superposition of a series of one-dimensional XRD scans. They are usually represented in a two-dimensional ($Q_x$, $Q_z$) plane, where $Q_x$ and $Q_z$ are the components of the scattering vector **Q**, respectively parallel and normal to the crystal surface, and they correspond to a section of the reciprocal space along this particular plane. Any structure obtained from an atomic scale simulation can be transformed into a RSM via the following equation:

$$I(Q_x, Q_z) = \sum_{\Delta Q_y} \left| \sum_{j=1}^{N} f_j(\mathbf{Q}) \exp\left(i\mathbf{Q} \cdot \mathbf{r}'_j\right) \right|^2 \quad (1)$$

where $\mathbf{r}'_j$ and $f_j$ are the coordinate vector of the $j^{th}$ atom and the atomic scattering factor of $j^{th}$ atom in the simulation cell, respectively. $Q_{x,y,z}$ are the three components of the scattering vector **Q**:

$$\mathbf{Q} = 2\pi\left(H\mathbf{a}^* + K\mathbf{b}^* + L\mathbf{c}^*\right) \quad (2)$$

where $\mathbf{a}^*, \mathbf{b}^*, \mathbf{c}^*$ are the reciprocal space basis vectors and $H$, $K$, $L$ are the so-called continuous Miller indices that determine the magnitude and direction of the scattering vector. Integer values of $H$, $K$ and $L$ correspond to maxima of the intensity distribution, *i.e.* Bragg peaks. Selecting different *HKL* values allows one to probe different directions of the simulation cell, *via* the scalar product $\mathbf{Q} \cdot \mathbf{r}'_j$ in Eq. (1). Finally $\Delta Q_y$ is a narrow integration range perpendicular to the ($Q_x$, $Q_z$) plane aimed at smoothing the computed RSM (see Appendix A).

The structural information that can be accessed *via* the RSMs is summarized in the following equation (Holý *et al.*, 1999):

$$I(\mathbf{Q}) = \sum_j \sum_k f_j(\mathbf{Q}) f_k^*(\mathbf{Q}) \times \langle \exp\{i\mathbf{Q}[\delta\mathbf{u}(\mathbf{r}_j) - \delta\mathbf{u}(\mathbf{r}_k)]\}\rangle$$
$$\times \exp(i\{\mathbf{Q}^\mathbf{T}[\mathbf{I} + \nabla\mathbf{u}(\mathbf{r}_j)]\} \cdot [\mathbf{r}_j - \mathbf{r}_k]) \quad (3)$$

where **r** are the atomic coordinates in the unperturbed lattice, and **u(r)** are the deviations from the perfect lattice, *i.e.* the atomic displacement vectors, so that $\mathbf{r}' = \mathbf{r} + \mathbf{u}(\mathbf{r})$. The displacement **u(r)** can be further separated into two components, $\mathbf{u}(\mathbf{r}) = \nabla\mathbf{u}(\mathbf{r}) \cdot \mathbf{r} + \delta\mathbf{u}(\mathbf{r})$, where $\nabla\mathbf{u}(\mathbf{r})\cdot\mathbf{r}$ describes the average response of the lattice to the presence of crystal defects. This term is a Taylor series expansion of the non-random displacement **u**, limited to the first order, which implicitly implies that the component of $\nabla \mathbf{u}(\mathbf{r})$ are much smaller than 1 (small strain approximation). In the following, $\nabla \mathbf{u}$ is denoted as the *homogeneous strain* tensor[1]. The second term, $\delta \mathbf{u}(\mathbf{r})$, corresponds to random local deviations around the average displacement.

The first exponential on the right-hand side of Eq. (3) contains the above-mentioned random displacement term and gives rise to two distinct effects. The first effect is a lowering of the Bragg (coherent) diffracted intensity quantified by the so-called static Debye-Waller (DW) factor (Warren, 1969; Krivoglaz, 1996). For a perfect crystal, $\delta \mathbf{u}(\mathbf{r}) = 0$, so that DW = 1 and the coherent intensity has its maximum value. On the contrary, for highly disordered crystals, or amorphized materials, $\delta \mathbf{u}(\mathbf{r}) \gg 0$, so that DW $\rightarrow$ 0 and the coherent intensity vanishes. The intensity subtracted from the Bragg scattering is redistributed in the background with a reciprocal space distribution depending on the degree of correlation of the disorder (*via* the $\delta \mathbf{u}(\mathbf{r}_j) - \delta \mathbf{u}(\mathbf{r}_k)$ term): this second effect is referred to as diffuse scattering. It is recalled in Appendix B that, in the case of correlated displacements, diffuse scattering manifests itself as peak broadening which can be connected to the rms strain tensor, *i.e.* the *heterogeneous strain* tensor.

The second exponential contains the effect of homogeneous (long-range) strain. The diffraction condition implies that the argument of this exponential be an integer multiple of $2\pi$. If, in a strain-free crystal, diffraction occurs at $\mathbf{Q}_0$, the presence of homogeneous strain, *via* $(\mathbf{I} + \nabla \mathbf{u})$, shifts the Bragg peak from its strain-free position $\mathbf{Q}_0$, to a modified coordinate $\mathbf{Q}_0^T \cdot (\mathbf{I} + \nabla \mathbf{u})^{-1}$, which, in the limit of small strain reduces to $\mathbf{Q}_0^T \cdot (\mathbf{I} - \nabla \mathbf{u})$. Depending on the direction of $\mathbf{Q}_0$, different components of $\nabla \mathbf{u}$ can be selected. For instance, a $\mathbf{Q}_0 = (0, 0, Q_z)^T$ vector is transformed into $(-e_{zx} Q_z, -e_{zy} Q_z, (1 - e_{zz})Q_z)^T$, where $e_{ij}$ are the components of $\nabla \mathbf{u}$. Measuring the coordinates of the Bragg peak for different $\mathbf{Q}_0$ values, hence, in principle, allows to retrieve the complete strain tensor.

**3. Computational details**

**3.1. Molecular dynamics**

MD simulations were performed using the Frenkel pairs accumulation (FPA) methodology (Chartier *et al.,* 2005; Crocombette *et al.,* 2006) in order to mimic ion-irradiation induced ballistic damages. This methodology avoids the calculation of complete displacement cascades and their accumulation by directly creating their final states, *i.e.*, point defects. It has been proven very efficient to simulate irradiation damages in different oxides (Chartier *et al.,* 2009; Catillon & Chartier, 2014) and graphite (Chartier *et al.,* 2018). In MD simulations, only uranium Frenkel pairs were created, which imply to measure the introduced disorder in displacement per uranium (dpU). MD calculations were performed in a 26 × 26 × 26 nm$^3$ UO$_2$ cell (768 000 atoms) using a Morelon empirical potential (Morelon *et al.,* 2003) which exhibits a relevant responses to irradiations (Devanathan *et al.,* 2010). FPA was performed at 0 pressure and at 300 K using a modified version of the large-scale atomic/molecular massively parallel simulator (LAMMPS) code (Plimpton, 1995) by creating 800 uranium Frenkel pairs(Crocombette & Chartier, 2007) every 2 ps. During the simulation, 898 snapshots of the MD cell, covering a 0 – 7.85 dpU range, were saved for further processing.

---

[1] Strictly speaking , $\nabla \mathbf{u}$ is the Jacobian of the displacement (the displacement gradient tensor), with components $e_{ij} = \partial u_i / \partial j$. It is here denoted as the strain tensor, although it is important to note that, in contrast to the actual strain tensor, $(\nabla \mathbf{u} + \nabla \mathbf{u}^T) / 2$, the displacement gradient tensor includes rigid rotations (which are important parameters to derive from the MD cells).

In addition, we created cells with rotated crystallites contained 768000 atoms with the perfect fluorite structure. Each cell was divided into 8 identical cubic sub-cells, half of which were rotated by 2° along an axis defined by a vector **r**$_{x,y,z}$, where the subscript indicate its orientation with respect to the cell (see also Fig. 2). We considered reflections with *HKL* = 002 and 004, and a rotation axis directed along the **x**, **y** and **z** direction of the cell.

**3.2. Computation and analysis of the reciprocal space maps**

RSMs were computed using Eq. (1). In the presence of disorder, the atomic coordinates can not be represented on a regular grid, which prohibits the use of efficient fast Fourier transform algorithms. For a typical RSM, as those represented in Fig. 2 for instance, the direct evaluation of Eq. (1) using a naive implementation requires more than one hour to reach completion which is incompatible with the large number of MD cells to be analyzed. Equation (1) was therefore evaluated on a graphics processing unit (GPU) using the Python programming language together with the NumPy (van der Walt *et al.,* 2011) and the PyNX (Favre-Nicolin *et al.,* 2011) libraries. With the hardware used in this work (Nvidia Quadro P5000 with a theoretical 64bits peak performance of 277 GFLOPS) the computation time drops to ~20 seconds for a single RSM. Because of the large number of snapshots to be analyzed, the data processing (including peak finding, peak fitting, etc.) is entirely automated using Python scripting as described below.

RSMs for *HKL* = 002 and 004 have been computed for 6 different orientations of the MD cell, that is with the [100], [010], [001] directions successively set parallel to **Q** vector and, for each direction, two 90° spaced azimuthal orientations have been considered. Doing so, for each *HKL*, all 6 off-diagonal components of the strain can be determined, and the 3 diagonal components (normal strains) are determined twice.

Each computed RSM was processed as follows:

- A peak finding algorithm detects all the maxima in the RSM and the corresponding ($H$, $L$) coordinates are saved. The peaks are indicated as circles in Fig. 3. In order to avoid the detection of minor peaks or interference fringes, a detection threshold of 30% of the maximum intensity has been defined. There is no scientific justification behind this value; it has been empirically determined so as to be low enough to record all high intensity peaks, and high enough in order to reject minor peaks. Lowering this value does not change the conclusions but complicates the analysis with the occurrence of several low intensity peaks. On the other hand, increasing this value too much may lead to erroneous conclusions regarding the development of strains, since high intensity peaks might be missed. These coordinates are then used to compute the diagonal and off-diagonal components of the strain tensor, *i.e.*

$$e_{ii} = \Delta L/L_0 \text{ and } e_{ij} = \Delta H/L_0 \tag{4}$$

where *i, j = x, y, z* are determined by the orientation of the MD cell. Where the subscript 0 indicates the virgin cell (0 dpU).

- For the maximum intensity peak, the integrated (coherent) intensity is computed by integrating the intensity along the row $H = H_{max}$, where $H_{max}$ is the $H$ coordinate of the most intense peak. For *HKL* = 002 and 004, and for weak disorder, $H_{max} = 0$, whereas deviations from $H_{max} = 0$ indicate rotations or shear strain, as discussed above. The integrated intensities are then used to compute the DW factor, which, correspond to the lowering of the coherently scattered amplitude relative to the perfect crystal case that is:

$$DW = \sqrt{\frac{I^{H_{max}}}{I_0^{H_{max}}}} \frac{v}{v_0} \tag{5}$$

where *v* is the volume of the MD cell. The appearance of the MD cell volume stems from the fact that the scattered x-ray amplitude is proportional to the materials density so that an increase of the cell volume with a fixed number of atoms yields a dilution of the scattered amplitude, resulting in fictitious drop of DW even without defects being introduced.

- The envelope of the intensity distribution was fitted with a bivariate asymmetrical Gaussian distribution (dotted contour lines in Fig. 3):

$$g^{(q,r)}(H, L) = I_{max} \exp\left\{-4\ln 2 \left[\frac{(H - H_{max})^2}{w_H^{(q,r)2}} + \frac{(L - L_{max})^2}{w_L^{(q,r)2}}\right]\right\} \tag{6}$$

where $w_{H,L}$ are the full-widths at half-maximum (FWHM) in the $H$ and $L$ directions and the superscripts *q* and *r* designate the lower and upper half of the width, that is:

$$w_{H,L}^{(q,r)} = \begin{cases} w_{H,L}^{(q)} & \text{if } H \leq H_{max} \text{ or } L \leq L_{max} \\ w_{H,L}^{(r)} & \text{otherwise.} \end{cases} \qquad (7)$$

The FWHM in the *H* and *L* directions is then obtained from:

$$w_{H,L} = (w_{H,L}^{(q)} + w_{H,L}^{(r)})/2 \qquad (8)$$

Notice that we here used a definition of the Gaussian distribution using FWHMs instead of standard deviations. Both are related *via* a $2\sqrt{2\ln 2}$ factor (see below). The corresponding FWHMs are used to determine the rms strain and shear/rotations.

The FWHMs determined from the fitting procedure contain both the contribution of defects and the contribution of the finite cell size. In order to extract the sole contribution of defects broadening, the cell size effect must be deconvoluted from the total FWHM which, in the case of Gaussian function, is straightforwardly performed:

$$w_{defects} = \sqrt{w^2 - w_{size}^2} \qquad (9)$$

where $w_{\text{size}}$ is given by the FWHM at low disorder levels. In the situation where defects only induce strain (as is the case in the present study) the diagonal and off-diagonal components of rms strain tensor are :

$$\epsilon_{i,j} = w_{H,defects}/(2\sqrt{2\ln 2}L) \qquad (10)$$

$$\epsilon_{i,i} = w_{L,defects}/(2\sqrt{2\ln 2}L) \qquad (11)$$

## 4. Results and discussion

### 4.1. Rotated crystalline domains

Before addressing the case of disordered $UO_2$ cells, we shall first consider a simpler example consisting of $UO_2$ cells containing crystalline domains half of which are rotated by 2° along the *x*, *y* or *z* axes. This is shown in Fig. 2g-i, respectively. For simplicity, a simple cubic lattice is depicted. We considered 002 and 004 reflections, for which the **Q** vector can be written $(0, 0, Q_z)^T$. Under the action of each of the three rotations, it gets transformed into $[0, -Q_z \times \sin \omega, Q_z \times (1-\cos \omega)]^T$, $[Q_z \times \sin \omega, 0, Q_z \times (1-\cos \omega)]^T$ and $(0, 0, Q_z)^T$, respectively, ω being the rotation angle. The corresponding RSMs are displayed in Fig. 2. The RSMs are plotted as a function of $\Delta H = H - H_0$ and $\Delta L = L - L_0$, which are the deviations from the unperturbed crystal indices, $H_0$ and $L_0$.

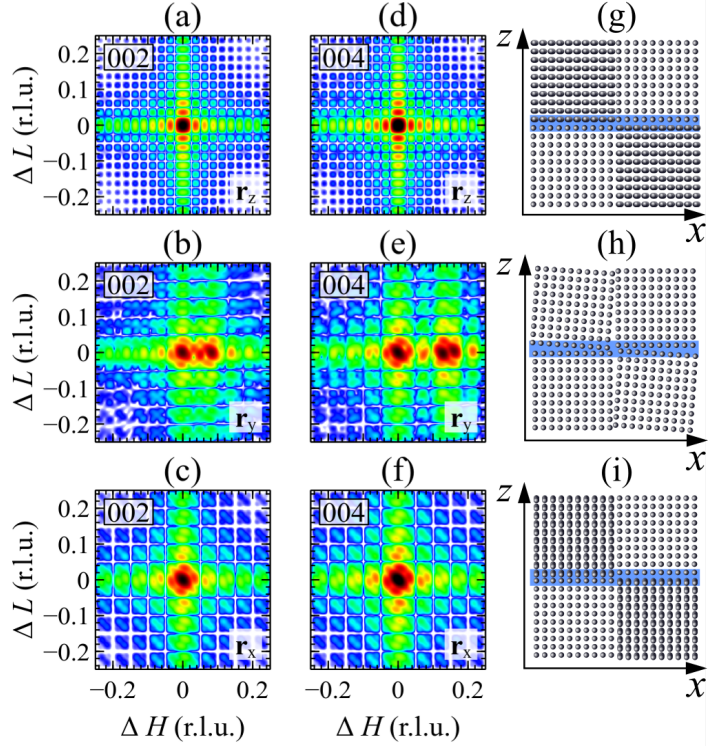

*Fig. 2. RSMs of $UO_2$ cells divided into 8 rigid sub-cells where half of the sub-cells are rotated by 2°, with a rotation axis directed along the z (a,d), y (b,e) and x (c,f) direction. (a,b,c) : 002 reflection. (d,e,f): 004 reflection. The intensity is plotted on a logarithmic scale using a usual red-yellow-green-blue color scale. The axes are graduated in reciprocal lattice units. The corresponding schematic atomic structures are given in (g,h,i).*

Let us consider the different cases in Fig. 2:
- Figures 2a,d correspond to the case $\mathbf{r}_z$ (rotation about the *z* axis). As indicated from the transformed **Q** vectors above, no modification is expected in this case and both RSMs correspond to the scattering from a perfect cubic crystal (a 2D Laue function). The interference fringes are related to the finite size of the MD cell and their period is given by the inverse of the number of $UO_2$ unit-cells in each direction, $1 / N_{uc} = 1 / 40 = 0.025$.

- Figures 2b,e correspond to the case $\mathbf{r}_y$. Since $\omega = 2°$ (0.035 rad), the transformed $\mathbf{Q}$ vector is close to $(- Q_z \times \omega, 0, Q_z)^T$, *i.e.* the crystallites affected by the rotation exhibit a Bragg peak shifted along the *H* direction, whereas the *L* coordinate remains unchanged. The result is a splitting of the Bragg peak, the magnitude of which being directly related to the rotation angle. The split for the 002 and 004 reflections is 0.07 and 0.14, respectively, consistently with a 2° rotation (*i.e.* 0.035×2 and 0.035×2). Another visible feature is the doubling of the fringe period and a broadening of the peak, as compared to Fig. 2a,b. This is due to the fact that the sub-cells are rigidly rotated and the interfaces between pristine and rotated sub-cells are incoherent and act as grain boundaries. For each sub-cell the summation in Eq. (1) is therefore truncated at *N*/2, hence the period doubling and the corresponding broadening (which is quantified by Scherrer's equation in the field of powder diffraction (Warren, 1969)). It can further be noticed that, contrarily to strain or rotation effects, this "finite crystallite size" effect is independent on the magnitude of $\mathbf{Q}$. This feature (*Q*-dependent *vs.* *Q*-independent effect) constitutes the basis of the size/strain separation in powder XRD line profile analysis methods (e.g. Williamson-Hall and Warren-Averbach methods (Warren, 1969)).

- Figures 2c,f correspond to the case $\mathbf{r}_x$. As for the previous case, the crystallites affected by the rotation have a $\mathbf{Q}$ vector close to $(0, -Q_z \times \omega, Q_z)^T$. However, contrarily to the previous case, because the RSMs result from an integration along the *y*-direction, they are not sensitive to the change in the $Q_y$ component, therefore no splitting is observed. Nonetheless, an indirect indication of the presence of this rotation is the fact that, as for $\mathbf{r}_y$, the fringes period is doubled and the peak width is increased, indicating the formation of grain boundaries.

An alternate way of picturing the effect of crystallite rotations on the RSMs is to observe the crystal structure projected in the (*x*, *z*) planes. Since the RSMs are computed with a $(0, 0, Q_z)^T$ vector, according to Eq. (3) we are solely sensitive to the *z* component of the atomic displacements:

- Figure 2g shows that when the rotation axis is parallel to *z*, no rotation nor any discontinuity in the atomic *z* coordinates are observed (see colored regions), hence the RSMs exhibiting a characteristic perfect 2D Laue intensity distribution (Fig. 2a,d).

- Figure 2h shows the effect of $\mathbf{r}_y$ (normal to the plane of the figure). Here, both finite size and rotation effects are clearly observed in the projected structure and in the RSMs (fig. 2b,e).
- Finally, Fig. 2i shows that, in the case of a $\mathbf{r}_x$ rotation axis (*i.e.* parallel to *x*), the atomic coordinates along the *z* axis exhibit abrupt variations both along the *x* and *z* axis (colored region), hence the observed finite size effect in the RSMs. On the contrary, no rotation is observed, in agreement with Fig. 2c,f.

This simple examples reveals that a RSM computed for a single 00*L* reflection allows to straightforwardly derive the $e_{zx}$ component of the strain tensor using Eq. (4). The other $e_{ij}$ components either have indirect effects (peak broadening and increased fringe spacing) or no effect at all. Although it is not explicitly illustrated in the example above, the $e_{zz}$ component is readily obtained by measuring the displacement of the Bragg peak along the $Q_z$ direction (as evident from Eq. (4)).

Therefore, using 6 independent *HKL* reflections or, equivalently, 6 orientations of the cell allows to retrieve the full strain tensor. In actual MD cells, the different components of the strain tensor occur simultaneously, with randomly oriented shear/rotations axis and with random distributions of normal strain and shear/rotation angles. The next sections illustrates how these different effects can be disentangled so as to determine the homogeneous strain tensor, the heterogeneous strain tensor, the disorder, etc.

## 4.2. Defective $UO_2$ MD cells

It is now well documented that the disordering kinetics of $UO_2$ submitted to ion-irradiation can be divided in different regions corresponding to a different dominant defects in the MD cell (Chartier *et al.,* 2016; Jin *et al.,* 2020), namely Frenkel pairs (from 0 to 0.09 dpU), Frank loops (from 0.09 to 0.3 dpU), perfect loops (from 0.3 to ~1.5 dpU) and dislocation lines (above ~1.5 dpU), respectively. In these studies, it was also demonstrated that the different types of defects can be correlated with the evolution of both the average elongation strain and the average disorder. In the following, we apply the computational diffraction approach to these $UO_2$ cells and proceed to the determination of the complete homogeneous and heterogeneous strain tensors, as well as the identification of strained/rotated crystalline domains.

We computed RSMs for *HKL* = 200, 020, 002, 400, 040 and 004, thereby setting the **Q** vector parallel to the **x**, **y** and **z** directions of the MD cell. Moreover, for each direction, two 90° spaced orientations around **Q** have been considered. RSMs of the 002 reflection at 4 selected disorder levels (0, 0.1, 0.3 and 3 dpU) are displayed in Fig. 3. These RSMs correspond to the MD cells shown in Fig. 1. However, contrarily to Fig. 1 where it is difficult to get clear insights regarding the defects and disorder affecting the cell, Fig. 3 shows features that are readily interpreted. The virgin cell corresponds to a 2D Laue function characteristic of a perfect crystal (Fig. 3a). At 0.1 dpU, Fig. 3b, we observe a displacement of the Bragg peak towards lower $L$ values which, according to the previous section, is indicative of the development of tensile strain. At this damage level, it can be observed that the intensity spreads out of the specular direction characterized by $H = 0$, a feature that is characteristic of diffuse scattering associated with atomic disorder. However, the coherent peak remains perfectly visible and aligned along the $H = 0$ row, which indicates that the level of disorder is low, and there is no evidence of rotation or shear strain. At 0.3 dpU, the coherent peak disappears and is replaced with a broad diffuse scattering peak, within which 3 peaks (*i.e.* 3 crystalline domains) can be detected; this corresponds to the situation where the highest levels of disorder and strain are present. Finally, at 3 dpU, the diffuse scattering coalesces back around the $H = 0$ coordinate, only one peak remains and interference fringes are reformed, which is indicative of a decrease in the level of disorder. The vertical shift of the reflection is also reduced, which demonstrates that the level of strain decreases as well. The Bragg peak is nonetheless broader than at 0 and 0.1 dpU and exhibits a fine structure, which may indicate the formation of rotated or sheared crystallites. In the following, we analyze the peak position, intensity and width in terms of strain, disorder and crystalline domains, respectively.

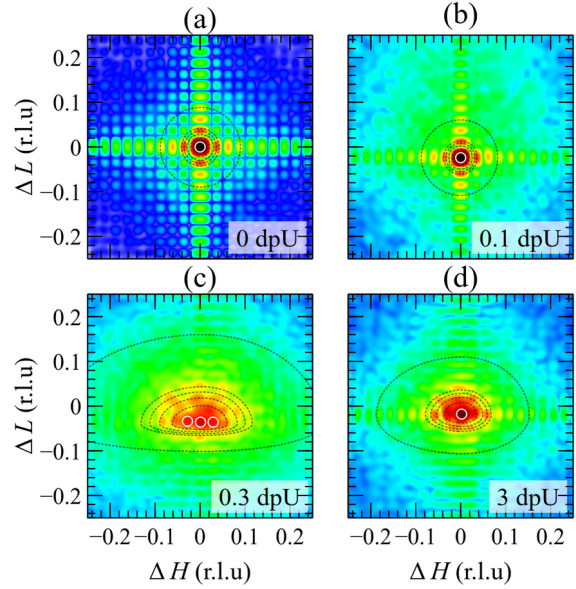

*Fig. 3. RSMs of the 002 reflection of UO$_2$ MD cells for disorder levels equal to 0 (a), 0.1 (b), 0.3 (c) and 3 (d) dpU. These dpU levels correspond to boundaries between different regions of the damage build-up kinetic. The intensity is plotted on a logarithmic scale using a red-yellow-green-blue color scale. Axes are graduated in reciprocal lattice units. White circles indicate the location of the peak maxima, and dotted lines are iso-contour lines of the 2D fit with asymmetric Gaussian functions.*

*Normal strains and disorder*

The normal strains (*i.e.* the diagonal elements of the homogeneous strain tensor) and the DW factor for the 6 different 002 RSMs are given in Fig. 4 for different dpU levels. Let us first consider a single orientation, say, Fig. 4a. The circles represent the level of strain, deduced from the vertical peak coordinates. The symbol color indicates the intensity ratio of the peak from which the strain is determined relatively to the virgin crystal, using a usual red-yellow-green-blue color scale: a high intensity (red) corresponds to a large crystal with low disorder, whereas a low intensity (blue) corresponds to small crystallites or high disorder. The red line is the DW factor which, by definition, is the disorder-induced intensity lowering and is therefore completely correlated with the intensity ratio given by the circles color. The background colors represent different steps of the UO$_2$ disordering kinetics, corresponding to the different defect stage mentioned above.

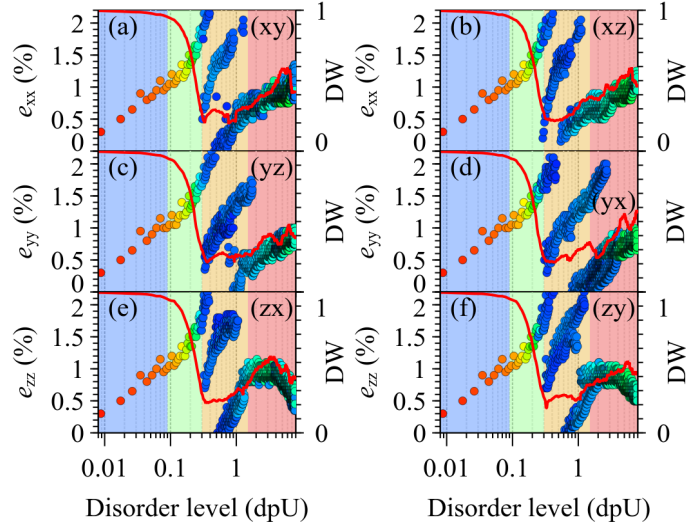

*Fig. 4. Evolution of the strain (filled circles, left axis) and Debye-Waller factor (red line, right axis) with increasing dpU. The symbol color is proportional to the intensity ratio relatively to the virgin crystal. The different panels correspond to different orientations of the MD cell. (a,b): the x axis is set parallel to **Q**. (c,d): the y axis is set parallel to **Q**. (e,f): the z axis is set parallel to **Q**. The intensity is integrated along z (a,d), y (b, e) and x (c, f). The background colors correspond to the different stages discussed in the text.*

The first region (0 to 0.09 dpU, light blue) is characterized by a steep increase of the strain and a low level of disorder. The second region (0.09 to 0.3 dpU, light green) shows a continuous increase of the strain[2] and a considerable increase in disorder, as attested by the drop of the DW factor. In the third region (0.3 to ~1.5 dpU, light orange), the DW factor is plateauing whereas the strain exhibits a very perturbed behavior: for a given disorder level, several crystalline domains (with different levels of strain) are observed, giving rise to Bragg peaks splitted from the main peak (as illustrated in the previous section). In this region, the strain decreases from ~2% at 0.3 dpU to ~0.5% at 1.5 dpU. However, this decrease takes place abruptly at discrete dpU values. Concomitantly, between two successive strain drops, strain accumulation continues as the consequence of the continuing disorder introduction. Finally, in the last region (above ~ 1.5 dpU), we observe a slight apparent healing of the material, as indicated by the increase of the DW factor. In this particular case the strain increases but, as further discussed below, its evolution depends on the orientation of the MD cell. Moreover, the exact extent of this last region also depends on the orientation.

---

2   Although this is not obvious from the figure, because of the horizontal logarithmic scale, the slope in the first region is ~8.9%/dpU *vs.* ~5%/dpU. The strain build-up kinetic is reduced in region 2.

Fig. 4 reveals that the first two steps take place identically, irrespective of the orientation of the MD cell. However, within stage 3, the number of observed strained regions, their level of strain as well as the disorder level required to trigger the strain drop strongly depends on the orientation of the MD cell. This observation points to a Frank → perfect dislocation loop reaction mechanism (Chartier *et al.,* 2016; Jin *et al.,* 2020), which, every time it takes places, locally lowers (*i.e.* relaxes) the strain in the region surrounding the initial Frank loop, hence the abrupt character. Moreover, as mentioned above, even in partially relaxed crystallites, the level of strain continues to increase (up to ~ 1.5 dpU when complete relaxation takes place) as a consequence of the continuous disordering. Above ~1.5 dpU, the strain exhibits an anisotropic behavior: whereas it increases along the [100] (Fig. 4a,b) and [010] directions (Fig. 4c,d), it decreases along [001] (Fig. 4e,f). This is due to the fact that, at high dpU, dislocations form a tangled network (Chartier *et al.,* 2016) that leads to the formation of sub-grain boundaries (see below). These boundaries act as traps for the interstitials, and the trapping efficiency depends on the number of boundaries (projected on a 2D plane) in the probed direction. The disorder also exhibits an anisotropic behavior: on average, it decreases for all orientations, but for the [100] and [001] directions an increase is observed. It should here be reminded that during the MD simulations, the cells are allowed to swell or shrink in all three directions, so that the observed anisotropy is not related to dimensional constraints imposed on the cells.

*Rotations, shear strains and crystalline domains*

Similarly to the normal strains, the rotation and shear strains can be straightforwardly determined from the position of the peaks visible in the RSMs. Figure 5 shows the evolution with disorder level of the off-diagonal elements of the strain tensor determined from the 002 RSMs (together with the DW factor, for comparison purposes). Up to 0.3 dpU, all components are equal to 0, indicating the absence of shear strains or rotations. However, as soon as the Frank loops start to transform (in the third stage, above 0.3 dpU), a significant disorder is introduced as evidenced by the appearance of low intensity peaks corresponding to several crystalline domains with distinct strain/rotation values randomly distributed in a ±1.5° range. These crystalline domains exactly correspond to those already observed in the case of the normal strains, but they are more clearly distinguished here because the difference in shear/rotations is more pronounced than the differences in the normal strains. As for the normal strains, the correlation with the DW factor is evident.

Although the disorder decreases in the last stage, some regions of the MD cells remain permanently rotated or sheared with seemingly randomly distributed values (0.5, 0.6, 0.8 and up to 1.5°). These random values are most likely due to the fact that the dislocations are randomly distributed within the MD cell. The number of sheared/rotated individual crystallites is in the 1-3 range depending on the orientation of the crystal with respect to **Q**. It can also be observed that in the $(x, y)$ and the $(y, z)$ planes, the disorder is mainly of rotational type, since the largest crystallites (indicated by the arrows in Fig. 5) are characterized by $e_{xy} = -e_{yx}$ and $e_{yz} = -e_{zy}$. On the contrary, in the $(x, z)$ plane the crystallites exhibit shear strain, with $e_{xz} = e_{zx}$.

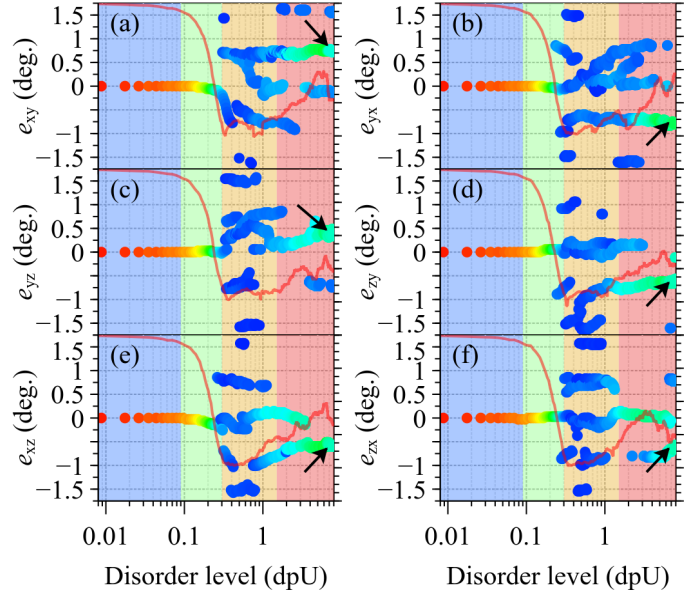

*Fig. 5. Evolution of the different components of the strain tensor as a function of the disorder level. The symbol color is proportional to the intensity ratio relatively to the virgin crystal. (a, b): the x and y directions are respectively set parallel to **Q**, while the intensity is integrated along z. (c,d): the y and z directions are respectively set parallel to **Q**, while the intensity is integrated along x. (e,f): the x and z directions are respectively set parallel to **Q**, while the intensity is integrated along y. The background colors correspond to the different stages discussed in the text. The thin red line is the DW factor (right axis, graduated from 0 to 1).*

These results show that the dislocation network that forms in the late stages of the disordering process yields to the formation of sub-grain boundaries rotated and sheared with respect to each other. However, contrarily to the case discussed in Fig. 2, where abrupt and incoherent interfaces were formed between the crystallites, eventually leading to finite-size broadening, there is no indication of such boundaries here, *i.e.* the transition from one rotated individual to another is continuous and there is no finite-size induced loss of coherence. This can be clearly confirmed from Fig. 6, which is a close-up of Fig. 1d taken at the intersection of several crystalline domains.

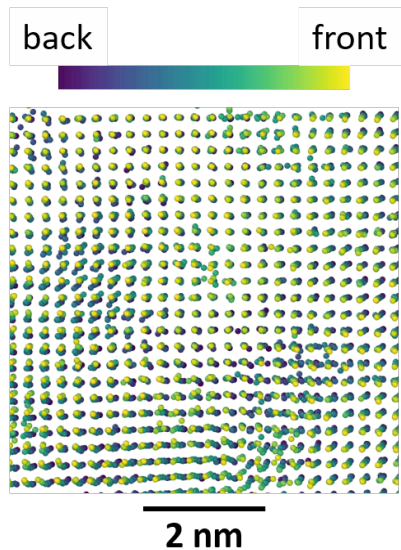

*Fig. 6. Close-up in the center of an MD cell at 3 dpU. The atomic displacements are continuous across different regions of the cells.*

*Towards larger cells: heterogeneous strains and rotations*

Because of the small size of the MD cells considered in this work ($26 \times 26 \times 26$ nm$^3$), individual crystalline domains can be detected as individual peaks in the RSMs. However, due to the random distribution of shear/rotation angles previously evidenced, it can be expected that in larger cells, the different peaks emanating from numerous domains, will eventually overlap to form a continuous intensity distribution. Although, in the present study, we do not have such large cells, an estimation of the corresponding effect can be inferred by analyzing the overall envelope of the RSMs (dotted contour lines in Fig. 3). Figures 7a-b show the evolution of the FWHM of the intensity distribution, averaged over all orientations, along the *H* and *L* directions, respectively, for both 002 and 004 reflections. The inset in Fig. 7a shows an example of a simulation of an intensity profile along *H* (corresponding to the 3 dpU case of Fig. 3) from which the FWHM is derived.

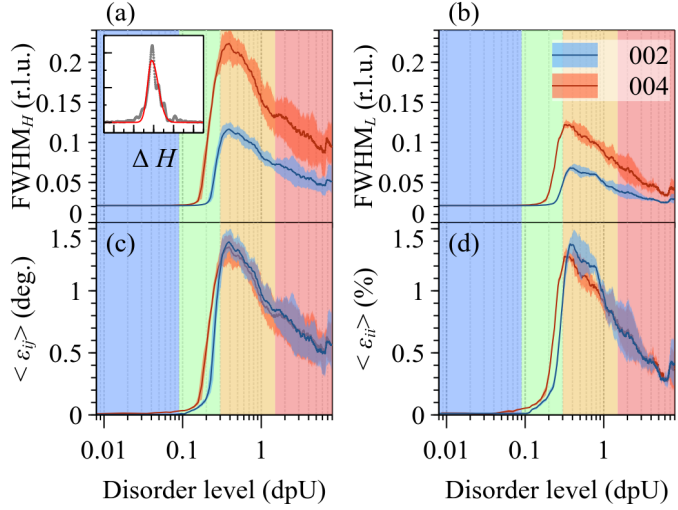

*Fig. 7. Evolution, for both the 002 and 004 reflections, of the FWHM along the H (a) and L (b) directions, and the corresponding rms shear/rotation (c) and rms strain (d). Inset: example fit of the intensity distribution with an asymmetric Gaussian function (grey points: data points; red line: fit). The horizontal axis extends from -0.25 to 0.25 r.l.u. The colored bands correspond to the uncertainty given by the minimum and maximum values observed for each disorder level. The background colors correspond to the different stages discussed in the text.*

The evolution of the FWHM is strikingly similar to the evolution of the DW factor. This is not unexpected since both parameters depend on the atomic displacement δ**u**(**r**) in the MD cell. Up to 0.1 dpU, the FWHM is constant and non zero because of the finite size of the MD cells. The FWHM reaches a maximum at the beginning of stage 3 and then steadily decreases. It can be observed that, starting from 0.3 dpU, the width obtained for the 002 and 004 reflections are within a ratio of 2, which is characteristic of a strain/rotation disorder (Boulle *et al.*, 2005). Below 0.3 dpU, the ratio differs from 2 because the overall FWHM results from the convolution of both the finite cell size (for which the FWHM ratio is 1) and the contribution of the disorder (for which the FWHM ratio is 2). This effect can be corrected by deconvolution (see section 3) and, as recalled in Appendix B, the components of the heterogeneous strain tensor, $\varepsilon_{ij}$, can be computed from the deconvoluted FWHM.

The evolution of $\varepsilon_{ij}$, averaged over all orientations, is shown in Fig. 7c,d. The values obtained for the 002 and 004 reflections are identical, within the numerical uncertainty. This finding shows that no finite size effect contributes to the overall broadening (apart from the MD cell size), *i.e.* no incoherent grain boundaries are formed and the atomic coordinates vary smoothly when moving to one rotated/sheared crystal to another, as inferred from Fig. 6. For the highest dpU, the residual $\varepsilon_{ij}$ is ~0.6°, which means that, with a normal distribution of rotations, the maximum deviation shall be around ±3 times the standard deviation, that is ±1.8°, which agrees very well with values observed in Fig. 5. Similarly, the maximum strain fluctuations should be around ~ 3 × 0.4 = 1.2%, which also agrees with the values observed in Fig. 4. This is results demonstrates that, although the MD cell has a limited spatial extension, the rms strain fluctuations that would be observed in a large crystal can be extrapolated by the analysis of the peak envelope.

**4.3. Real-space vs. reciprocal-space determinations**

An important question to address is how are these reciprocal-space – based strain measurements connect with more direct (*i.e.* real-space) determinations performed from the MD cells[3]. As an example, we shall consider here the results obtained for the average homogeneous strain $<e_{ii}>$ and the average heterogeneous strain $<\varepsilon_{ii}>$. The former is given by averaging the values displayed in Fig.4a-f, and the latter is given in Fig. 7d. At this point, the mathematical and numerical details of the methodology are beyond the scope of the paper, so we here only provide the most important results. The reader may refer to Appendix C and Supplementary material where all relevant details are given.

As mentioned in the Introduction, the local strain determined from the MD cell is sensitive to displacements affecting the first neighbors, or the atoms of the first unit-cell (Stukowski *et al.*, 2009; Xiong *et al.*, 2019). As a consequence, long-range, spatially correlated atomic displacements are not captured in this measurement, even if the local strain is averaged over the whole MD cell. To circumvent this issue, we introduce the 1D pair distribution function (1D-PDF), which correspond to the probability of finding a pair of atoms separated by a distance comprised between $z$ and $z + \Delta z$ along a given direction. In accordance with the previous section we here choose these directions to be [100], [010] and [001]. The 1D-PDF exhibits a series of peaks, the position of which corresponds to the $n^{th}$ neighbor distance, and the width of which corresponds to the rms fluctuations of this distance. The analysis of the 1D-PDF peak positions and widths, as function of the neighbor index $n$, therefore allows to determine the long-range strain in real-space, taking into account spatial correlations.

The corresponding results are shown in Fig. 8, for both $<e_{ii}>$ and $<\varepsilon_{ii}>$, on the one hand, from computational diffraction and, on the other hand, from real-space measurements considering, or disregarding, long-range correlations. For the homogeneous strain, (Fig. 8a), a remarkably good agreement is obtained between computational diffraction and the real-space measurement. It can also be noticed that when the strain is determined from the first neighbor only (which corresponds to a plain spatial average of the local strain), correct values are obtained for levels of strain lower than ~0.8%, *i.e.* when the dilatation of the unit-cell is solely induced by point defects (both at low and high dpU levels). As soon as spatial correlations are present, the strain computed from the first neighbor is incorrect by up to a factor 2.

---

[3] It must be noticed that these measurements differ from the "macroscopic" swelling of the MD cell, in particular as soon as the concentration of interstitials and vacancies are no longer equal.

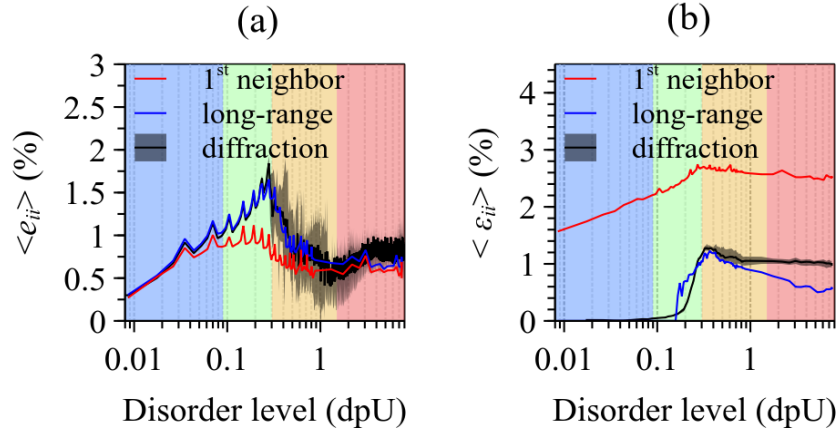

*Fig. 8. (a) evolution of the homogeneous strains determined from computational diffraction (black line), using the 1D-PDF (blue line) and using the first neighbor distance only (red line). The gray areas correspond to the uncertainty given by the minimum and maximum values observed for each disorder in the diffraction measurement. (b) same curves as in (a) but for the heterogeneous strains.*

The same type of analysis can be performed for the heterogeneous strain. Considering only the rms fluctuations of the local strain, *i.e.* the width of the first peak of the 1D-PDF, greatly overestimates the actual level of heterogeneous strain (Fig. 8b, red line). The reason for this failure is that the overall 1D-PDF peak broadening includes both correlated and uncorrelated disorder. Correctly taking into account spatial correlations (Fig. 8b, blue curve), via the introduction of a correlation length, $\xi$, above which atomic displacements cease to accumulate (Boulle *et al.,* 2005), allows to correctly determine the heterogeneous strain from a real-space analysis of the MD cells (see Appendix C). It can be observed that the agreement between the real-space and the reciprocal-space determinations is less perfect than for the homogeneous strain. The most likely reason for this discrepancy is that the real-space measurements require a more complex numerical analysis, which is more prone to numerical errors than computational diffraction.

Finally, the main interest of computational diffraction as compared to a real-space analysis, is that the former allows for a straightforward determination of the complete (homogeneous and heterogeneous) strain tensors. On the contrary, while the diagonal elements are accessible from a real-space measurement, as evidenced in this section, the off-diagonal elements would be far more complicated to assess.

## 5. Conclusions

We presented a computational diffraction method for an accurate determination of the homogeneous and heterogeneous (rms) strain tensors, as well as the associated disorder and crystalline domains, in atomic scale simulations of disordered or nanostructured crystals. The analysis is based on a projected 3D Fourier transform followed by a 2D peak position and shape analysis. With an adapted hardware, and depending on the size of the simulation cell, the whole process takes a few seconds for a given cell orientation. The main advantage of the computational diffraction approach, as compared to usual real-space approaches, is that the crystalline domains, homogeneous strain, heterogeneous strain and disorder (DW factor) are unambiguously and straightforwardly determined from the peak positions, widths and intensities in the reciprocal space maps. This means that the present approach is immune to issues associated with the definition of the local atomic strain in highly strained or damaged regions of the cell, or to the lack of clearly defined grain boundaries. Moreover, as compared to the Debye scattering equation, the present approach readily resolves individual crystalline domains, and the 3D strain tensor can be accurately determined without orientational averaging.

The relevance of computational diffraction has been demonstrated in a self-consistent way by comparing characteristic parameters (such as the normal homogeneous and heterogeneous strain) with those obtained from real space measurements. Since the whole analysis can be completely automated, it can easily be envisioned for on-the-fly data analysis where MD simulations snapshots are analyzed as soon as they are produced, allowing one to extract the data in real time without the need to save the simulation cells themselves. This *in situ* computational diffraction approach (similar to the *in situ* computational microscopy introduced in (Zepeda-Ruiz *et al.*, 2017)) allows one to significantly compress the amount of data to be stored.

Finally, although the efficiency of computational diffraction has been illustrated with the study of radiation damage, it can be used in any field where the microstructural characterization of MD cells is needed. Moreover, the approach is by no means restricted to MD simulations. Any atomic scale simulation, outputting a list of atomic coordinates, is compatible with the present approach, which potentially makes it appealing to the whole community of numerical scientists.

**Appendix A: Effect of $Q_y$ integration**

In most actual laboratory XRD experiments, x-ray beams are in general only collimated in one plane (the plane in which the photons are detected), whereas they are divergent in the orthogonal plane (this geometry is schematically depicted in Fig. S1, Supplementary material). This results in a loss of spatial coherence in the direction perpendicular to the detection plane and an intensity integration as given by Eq. (12), see also (Channagiri *et al.*, 2015).

In the present work we take advantage of this effect to approximate a Monte-Carlo averaging in order to wipe out interference features that are not usually observed in experiments:

$$\langle I(Q_x, Q_z) \rangle = \int dQ_y \cdot |E(\mathbf{Q})|^2 \approx \sum_{\Delta Q_y = -N_y/2}^{N_y/2} |E\left[(Q_x, Q_z)_{\Delta Q_y}\right]|^2 \qquad (12)$$

where $E$ is the usual scattered X-ray amplitude $\sum_{j=1}^{N} f_j(\mathbf{Q}) \exp(i\mathbf{Q} \cdot \mathbf{r}'_j)$, $N_y$ is the number of ($Q_x$, $Q_z$) sections involved in the summation, the appropriate value of which can be determined empirically by increasing $N_y$ to determine the the convergence of the sum (Fig. S2, Supplementary material). All calculations presented in the current work are computed using the configuration with 50 sections. An interesting analogy that can be made from this equation is that, in the diffraction experiment, the crystal can be viewed as divided into unit-cell-thick (*x, z*) planes, parallel to the detection plane. The intensity diffracted from each (*x, z*) plane is given by the squared modulus of the amplitude scattered from each plane, and the intensities are summed to form the total scattered intensity. This operation introduces randomness in an otherwise fixed defect configuration.

**Appendix B: Effect of heterogeneous strain**

The first exponential of Eq. 3 is also referred to as the correlation function (Holý *et al.*, 1999), G($\mathbf{r_j}$, $\mathbf{r_k}$), [or the strain Fourier coefficient (Warren, 1969) in the field of powder diffraction] and it determines the intensity and the shape of the diffuse scattering distribution. The important feature to notice is that G($\mathbf{r_j}$, $\mathbf{r_k}$) is a function of the *local displacement difference function* δ**u**($\mathbf{r_j}$) − δ**u**($\mathbf{r_k}$): it encodes the degree of correlation between local displacements, δ**u**, from one site to another (Boulle *et al.*, 2005). In the case of correlated disorder, the individual displacements add up, which yields a destruction of the long range order. In, such a case, G($\mathbf{r_j}$, $\mathbf{r_k}$) rapidly decreases for increasing $\mathbf{r_j} - \mathbf{r_k}$ pair distances (Krivoglaz, 1996). The displacement difference can therefore safely be expanded into a first order Taylor series so that the correlation function finally writes:

$$G(\mathbf{r_j}, \mathbf{r_k}) = \langle \exp\{i[\mathbf{Q}\nabla\delta\mathbf{u}(\mathbf{r})] \cdot (\mathbf{r_j} - \mathbf{r_k})\}\rangle \qquad (13)$$

$\nabla \delta \mathbf{u}(\mathbf{r})$ is the Jacobian of the local displacements. In order to illustrate how $G(\mathbf{r}_j, \mathbf{r}_k)$ affects the scattered intensity, let us consider the same two extreme cases as for the coherent scattering.

For a perfect crystal, $\delta \mathbf{u}(\mathbf{r}) = 0$ so that $G(\mathbf{r}_j, \mathbf{r}_k) = 0$, and there is no diffuse scattering, *i.e.* DW = 1. On the contrary for highly disordered crystals, $\delta \mathbf{u}(\mathbf{r}) \gg 0$, DW $\rightarrow$ 0, the coherent intensity vanishes and the shape of the intensity distribution is then entirely governed by the correlation function, which is itself a function of the statistical defect distribution and the associated displacement fields; increasing correlated disorder, for instance by increasing the defect density, yields a broadening of the intensity peak. An exact solution for the correlation can be worked out for selected defects such as point defects, defect clusters, dislocation loops,etc (Dederichs, 1971, 1973; Ehrhart *et al.,* 1982; Iida *et al.,* 1988; Larson, 2019) but this is not the topic of the present article.

For illustration purposes, we shall assume that the statistical distribution of the local displacement difference function obeys a multivariate Gaussian distribution. The correlation function then writes

$$\langle \exp\{i [\mathbf{Q} \nabla \delta \mathbf{u}(\mathbf{r})] \cdot (\mathbf{r_j} - \mathbf{r_k})\} \rangle = \exp\left[-\frac{1}{2}(\mathbf{r_j} - \mathbf{r_k})^\mathbf{T} \Sigma_{\mathbf{Q}\delta\mathbf{u}} (\mathbf{r_j} - \mathbf{r_k})\right] \quad (14)$$

where $\Sigma_{\mathbf{Q}\delta\mathbf{u}}$ is 3×3 tensor whose components are related to the *heterogeneous strain tensor*. For instance, using a $\mathbf{Q} = (0, 0, Q_z)^\mathrm{T}$ vector and assuming, that the displacements $\delta u_x$, $\delta u_y$ and $\delta u_z$ are statistically independent, the tensor writes:

$$\Sigma_{\mathbf{Q}\delta\mathbf{u}} = \begin{pmatrix} (\epsilon_{zx} Q_z)^2 & 0 & 0 \\ 0 & (\epsilon_{zy} Q_z)^2 & 0 \\ 0 & 0 & (\epsilon_{zz} Q_z)^2 \end{pmatrix} \quad (15)$$

in which case, Eq. (14) reduces to the product of three Gaussian functions with standard deviations $\varepsilon_{zx} Q_z$, $\varepsilon_{zy} Q_z$, $\varepsilon_{zz} Q_z$, where $\varepsilon_{i,j}$ are the components of the heterogeneous strain tensor. In other words, measuring the width of the peak in the RSM allows to determine the heterogeneous strain, which is the basis of all XRD line width analysis techniques related to Williamson & Hall pioneering work (Williamson & Hall, 1953). It is also clearly established that this approach is an oversimplification that relies on strong assumptions (Gaussian distribution of strain, statistically independent displacements). It nonetheless allows to get reasonable orders of magnitude.

**Appendix C: Real-space vs. reciprocal-space determinations of strain**

To address this task we first suggest to rewrite the equation of the total diffracted intensity (Eq. 3), by introducing the $n^\mathrm{th}$ neighbor index, $n = k - j$. Without loss of generality we shall focus on the $Q_z$ intensity distribution of reflection with $\mathbf{Q}_0 = (0, 0, Q_0)^\mathrm{T}$. The conclusions drawn remain perfectly valid for any reflection or any direction, although the derivation of the corresponding real space quantities (in particular the shear/rotational components of the strain tensors) would not be as straightforward. The corresponding intensity writes:

$$I(Q_z) = \sum_j \sum_n f_j(Q_z) f^*_{j+n}(Q_z) \times \langle \exp\{iQ_z[\delta u_z(z_j) - \delta u_z(z_{j+n})]\}\rangle \qquad (16)$$
$$\times \exp(i\{Q_z[1 + e_{zz}(z_j)]\} \cdot [z_j - z_{j+n}])$$

The term in the angular brackets can be reduced to

$$\langle \exp\{iQ_z[\delta u_z(z_j) - \delta u_z(z_{j+n})]\}\rangle = \exp\left[-\frac{1}{2}Q_z^2 \langle \delta u_z(z_j)\rangle^2\right] \qquad (17)$$

where we implicitly assumed a Gaussian distribution to perform the average; more general distributions can be considered (Boulle *et al.*, 2005; Boulle & Debelle, 2016), but this does not change the conclusions. The strain values, $e_{zz}$, and the associated rms strain, $\varepsilon_{zz}$, computed from the MD data using these equations exactly correspond to those given in Fig. 4 and Fig. 7.

As detailed in section 2, we write the *z* coordinate of all atoms as

$$z' = z + e_{zz} \cdot z + \delta u_z(z) \qquad (18)$$

and we introduce the 1D pair distribution function (1D-PDF), $p_{1D}(z)$, which corresponds to the probability of finding a pair of atoms separated by a distance comprised between *z* and *z* + Δ*z* along the [001] direction. The 1D-PDF can directly be evaluated from the MD cells (see below). This function exhibits maxima at the average $n^{th}$ neighbor distance (Supplementary material, Fig. S3a):

$$\langle D(n)\rangle = \langle (z'_j - z'_{j+n})\rangle_j \qquad (19)$$

where the average is taken over all atoms *j* in the cell. Using Eq. (18), the $n^{th}$ neighbor distance can be rewritten

$$\langle D(n)\rangle = n\langle d(1)\rangle(1 + e_{zz}) \qquad (20)$$

where *d*(1) is the first neighbor distance in a disorder-free MD cell. The lattice strain can hence be directly deduced from the slope of the peak positions in the 1D-PDF *vs.* the neighbor index *n* (Fig. S3b). Moreover, comparing Eq. (20) with the last exponential of Eq. (16) reveals that the strain measured from calculated intensity should be strictly equal to the strain computed from Eq. (20).

Similarly, the variance of the peaks of the 1D-PDF is:

$$\langle D(n)^2 \rangle = \langle (z'_j - z'_{j+n})^2 \rangle_j = \langle \delta u_z(z)^2 \rangle_j \qquad (21)$$

which exactly corresponds to the term contained in Eq. (17). Depending on the state of correlation of the displacements, different behavior are observed. If all displacements are independent and uncorrelated, the width of the peaks of the 1D-PDF is constant and equal to $\sqrt{\langle \delta u_z^2 \rangle}$. On the contrary if all displacements add up, $\delta u_z(z) = \sum_{\zeta \leq z} \delta u_\zeta$, then the width of peaks of the 1D-PDF increase as $n\sqrt{\langle \delta u_z^2 \rangle}$. Intermediate behaviors require the introduction of an additional quantity, the correlation length $\xi$, which defines the length up to which the displacement are correlated; for neighbors separated by distances larger than $\xi$, correlation is lost and the disorder saturates to a constant value, $\sigma_\infty$.

A phenomenological equation capturing the above-mentioned behavior has been proposed in (Boulle *et al.*, 2005):

$$\sigma_z(z) = \sigma_\infty \left\{ 1 - \exp\left[-\left(\frac{z}{\xi}\right)^{1/w}\right] \right\}^{H_u/w} \qquad (22)$$

where $w$ defines the width of the transition region between the correlated and the uncorrelated regimes, and $H_u$ (the Hurst exponent) defines how $\sigma_z(z)$ scales with $z$ in the correlated regime; for $H_u = 1$, the scaling is linear as described in the previous paragraph. Equipped with Eq. (22) it is possible to model the evolution of standard deviation of the peaks of the 1D-PDF as a function of the neighbor distance, $\sigma_z$ *vs.* $z$, and hence to retrieve the values $\sigma_\infty$ and $\xi$ (Supplementary material, Fig. S3a). From these values the rms strain, $\varepsilon_{zz}$, is obtained with $\sigma_\infty / \xi$ (Boulle *et al.,* 2005). The parameters $H_u$ and $w$ are also obtained from the simulation, but their interpretation is beyond the scope of this paper.

In order to obtain the 1D-PDF from the MD cell, we calculated the distribution of distances between neighbours considering for each atom a centred square bar oriented in the direction of interest. We chose square bars with a basal section 1 Å$^2$ and a length of 80 Å as it turned out to be sufficient to capture the main trends.

**References**

Boulle, A. & Debelle, A. (2016). *Phys Rev Lett*. **116**, 245501.

Boulle, A., Guinebretière, R. & Dauger, A. (2005). *J. Phys. Appl. Phys.* **38**, 3907–3920.

Catillon, G. & Chartier, A. (2014). *J. Appl. Phys.* **116**, 193502.

Channagiri, J., Boulle, A. & Debelle, A. (2015). *J Appl Cryst*. **48**, 252.

Chartier, A., Catillon, G. & Crocombette, J.-P. (2009). *Phys Rev Lett*. **102**, 155503.

Chartier, A., Meis, C., Crocombette, J.-P., Weber, W. J. & Corrales, L. R. (2005). *Phys. Rev. Lett.* **94**, 025505.

Chartier, A., Onofri, C., Brutzel, L. V., Sabathier, C., Dorosh, O. & Jagielski, J. (2016). *Appl Phys Lett*. 7.

Chartier, A., Van Brutzel, L. & Pageot, J. (2018). *Carbon*. **133**, 224–231.

Clausius, R. (1870). *Ann. Phys.* **217**, 124–130.

Costanzo, F., Gray, G. L. & Andia, P. C. (2005). *Int. J. Eng. Sci.* **43**, 533–555.

Crocombette, J.-P. & Chartier, A. (2007). *Nucl. Instrum. Methods Phys. Res. Sect. B Beam Interact. Mater. At.* **255**, 158–165.

Crocombette, J.-P., Chartier, A. & Weber, W. J. (2006). *Appl. Phys. Lett.* **88**, 051912.

Debelle, A., Boulle, A., Chartier, A., Gao, F. & Weber, W. J. (2014). *Phys Rev B*. **90**, 174112.

Dederichs, P. H. (1971). *Phys. Rev. B*. **4**, 1041–1050.

Dederichs, P. H. (1973). *J. Phys. F Met. Phys.* **3**, 471–496.

Devanathan, R., Brutzel, L. V., Chartier, A., Guéneau, C., Mattsson, A. E., Tikare, V., Bartel, T., Besmann, T., Stan, M. & Uffelen, P. V. (2010). *Energy Environ. Sci.* **3**, 1406–1426.

Ehrhart, P., Trinkaus, H. & Larson, B. C. (1982). *Phys. Rev. B*. **25**, 834–848.

Favre-Nicolin, V., Coraux, J., Richard, M.-I. & Renevier, H. (2011). *J. Appl. Crystallogr.* **44**, 635–640.

Gelisio, L. & Scardi, P. (2016). *Acta Crystallogr. Sect. Found. Adv.* **72**, 608–620.

Gullett, P. M., Horstemeyer, M. F., Baskes, M. I. & Fang, H. (2008). *Model. Simul. Mater. Sci. Eng.* **16**, 015001.

Holý, V., Pietsch, U. & Baumbach, T. (1999). *High Resolut. X-Ray Scatt. Multilayers Thin Films Springer Tracts Mod. Phys.* **149**,.

Iida, S., Larson, B. C. & Tischler, J. Z. (1988). *J. Mater. Res.* **3**, 267–273.

Jin, X., Boulle, A., Chartier, A., Crocombette, J.-P. & Debelle, A. (2020). *Acta Mater.* **201**, 63–71.

Krasheninnikov, A. V. & Nordlund, K. (2010). *J. Appl. Phys.* **107**, 071301.

Krivoglaz, M. A. (1996). *X-Ray Neutron Diffr. Nonideal Cryst.*

Larson, B. C. (2019). *Crystals.* **9**, 257.

Lin, Z. & Zhigilei, L. V. (2006). *Phys. Rev. B.* **73**, 184113.

Morelon, N.-D., Ghaleb, D., Delaye, J.-M. & Brutzel, L. V. (2003). *Philos. Mag.* **83**, 1533–1555.

Mott, P. H., Argon, A. S. & Suter, U. W. (1992). *J. Comput. Phys.* **101**, 140–150.

Nordlund, K. & Djurabekova, F. (2014). *J. Comput. Electron.* **13**, 122–141.

Plimpton, S. (1995). *J. Comput. Phys.* **117**, 1–19.

Soulié, A., Menut, D., Crocombette, J.-P., Chartier, A., Sellami, N., Sattonnay, G., Monnet, I. & Béchade, J.-L. (2016). *J. Nucl. Mater.* **480**, 314–322.

Stukowski, A. (2012). *Model. Simul. Mater. Sci. Eng.* **20**, 045021.

Stukowski, A. & Arsenlis, A. (2012). *Model. Simul. Mater. Sci. Eng.* **20**, 035012.

Stukowski, A., Bulatov, V. V. & Arsenlis, A. (2012). *Model. Simul. Mater. Sci. Eng.* **20**, 085007.

Stukowski, A., Markmann, J., Weissmüller, J. & Albe, K. (2009). *Acta Mater.* **57**, 1648–1654.

Tadmor, E. B. & Miller, R. E. (2011). Modeling Materials: Continuum, Atomistic and Multiscale Techniques Cambridge University Press.

van der Walt, S., Colbert, S. C. & Varoquaux, G. (2011). *Comput. Sci. Eng.* **13**, 22–30.

Warren, B. E. (1969). X-Ray Diffraction.

Williamson, G. K. & Hall, W. H. (1953). *Acta Metall.* **1**, 22–31.

Xiong, S., Lee, S.-Y. & Noyan, I. C. (2019). *J. Appl. Crystallogr.* **52**, 262–273.

Zepeda-Ruiz, L. A., Stukowski, A., Oppelstrup, T. & Bulatov, V. V. (2017). *Nature.* **550**, 492–495.

Zhang, L., Jasa, J., Gazonas, G., Jérusalem, A. & Negahban, M. (2015). *Comput. Methods Appl. Mech. Eng.* **283**, 1010–1031.

Zimmerman, J. A., Bammann, D. J. & Gao, H. (2009). *Int. J. Solids Struct.* **46**, 238–253.